\documentstyle[11pt,newpasp,twoside,epsf]{article}
\def\kpc{{\rm kpc}}
\def\dls{{D_{\rm LS}}}
\def\te{{t_{\rm E}}}
\def\re{{r_{\rm E}}}
\def\umin{u_{0}}

\def\days{{\rm days}}
\def\au{{\rm AU}}
\def\ob14{OGLE-1998-BUL-14}
\def\dos{D_{\rm S}}
\def\dls{D_{\rm LS}}
\def\dol{D_{\rm L}}
\def\thetae{\theta_{\rm E}}

\def\dchi{\Delta\chi^2}

\def\murel{\mu_{\rm rel}}
\def\Packy{Paczy\'nski~}
\def\amax{A_{\rm max}}
\def\msun{M_\odot}

\def\edcomment#1{\iffalse\marginpar{\raggedright\sl#1\/}\else\relax\fi}
\reversemarginpar

\begin{document}
\title{Microlensing Constraints on the Frequency of Jupiter Mass Planets}
\author{B. Scott Gaudi$^1$, M.D. Albrow$^2$, Jin H. An$^1$, J.-P. Beaulieu$^3$,
J.A.R. Caldwell$^4$, D.L. Depoy$^1$, M. Dominik$^5$, A. Gould$^1$, J. Greenhill$^6$,
K. Hill$^6$, S. Kane$^6$, R. Martin$^7$, J. Menzies$^4$, R.W. Pogge$^1$, K. Pollard$^8$,
P.D. Sackett$^5$, K.C. Sahu$^2$, P. Vermaak$^4$, R. Watson$^6$,
A. Williams$^7$ 
(The PLANET collaboration)}
\affil{$^1$ The Ohio State University}
\affil{$^2$ Space Telescope Science Institute}
\affil{$^3$ Institut d'Astrophysique de Paris}
\affil{$^4$ South African Astronomical Observatory}
\affil{$^5$ Kapteyn Astronomical Institute}
\affil{$^6$ University of Tasmania}
\affil{$^7$ Perth Observatory}
\affil{$^8$University of Canterbury}

\begin{abstract}
Microlensing is the only technique likely, within the next 5 years, to
constrain the frequency of Jupiter-analogs.  The PLANET collaboration
has monitored nearly 100 microlensing events of which more than 20
have sensitivity to the perturbations that would be caused by a
Jovian-mass companion to the primary lens. No clear signatures of such
planets have been detected.  These null results indicate that Jupiter
mass planets with separations of 1.5-3 AU occur in less than 1/3 of
systems.  A similar limit applies to planets of 3 Jupiter masses between
1-4 AU.  
\end{abstract}

\section{Introduction}

A Galactic microlensing event occurs when a massive, compact object (the lens) passes near
to our line-of-sight to a more distant star (the source).  If the
lens, observer, and source are perfectly aligned, then the lens images
the source into a ring, called the Einstein ring, which has angular radius
of\footnote{For the scaling relation on the far right of equations
(1), (2), and (3), we have
assumed $\dos=8~\kpc$ and $\dol=6.5~\kpc$, typical distances to the
lens and source for microlensing events toward the bulge.}
\begin{equation}
\thetae \equiv\left[{4GM \over c^2} {\dls \over \dol \, \dos}\right]^{1/2}
\sim 480 \, \mu{\rm as} \left({M \over M_{\odot}}\right)^{1/2},
\label{eqn:thetae}
\end{equation}
where $M$ is the mass of the lens, and $\dls$, $\dos$, $\dol$ are the
lens-source, observer-source, and observer-lens distances,
respectively.  This corresponds to a physical distance at the lens
plane of
\begin{equation}
\re = \thetae \dol \sim 3~\au \left({M \over 
M_{\odot}}\right)^{1/2}.
\label{eqn:re}
\end{equation}

If the lens is not perfectly aligned with the line-of-sight, then the
lens splits the source into two images.  The separation of these
images is ${\cal O}(\thetae)$ and hence unresolvable.  However, the
source is also magnified by the lens, by an amount that depends on the angular
separation between the lens and source in units of $\thetae$.  Since the lens, observer, and
source are all in relative motion, this magnification is a function of
time: a `microlensing event.'  The time scale for such an event is 
\begin{equation}
\te \equiv { \thetae  \over \murel} \sim 40~\days \left({M\over
M_{\odot}}\right)^{1/2},
\end{equation}
where $\murel$ is the relative lens-source proper motion.  

If the primary lens has a planetary companion, and the position of
this companion happens to be in the path of one of the two images
created during the primary event, then the planet will perturb the
light from this image, creating a deviation from the primary
light curve (see Figure 1).  The duration of this perturbation is $\sim
\sqrt{q} \te$, where $q$ is the mass ratio between the planet and
primary.  Hence, for a Jupiter/Sun mass ratio ($q\simeq10^{-3}$), the
perturbation time scale is ${\cal O}({\rm day})$.  These short-duration
deviations are the signatures of planets orbiting the primary lenses.
Note that since the perturbation time scale is considerably less than
$\te$, the majority of the light curve will be indistinguishable from a
single lens. 

Three parameters determine the magnitude of the perturbation, and
hence define the observables.  These are mass ratio $q$, the
instantaneous angular separation $d$ between the planet and primary
in units of $\re$, and the angle $\alpha$ between the projected planet/star axis and the
path of the source.  As $q$ decreases, the perturbation time scale
decreases, although the magnitude of the deviation does not
necessarily decrease.  Thus very small mass ratio planets ($q \la
10^{-5}$) can be
detected using microlensing, although the detection probability is small.
The lower limit to the detectable $q$ is
set practically by the sampling of the primary event, and ultimately
by the finite size of the source stars (Bennett \& Rhie 1996).  
A microlensing event is generally alerted only if the 
minimum angular impact parameter in units of $\thetae$ satisfies $\umin \leq
1$, which corresponds to image positions between $(0.6-1.6)\thetae$.
Since the planet must be near one of these images to create a
perturbation, microlensing is most sensitive to planets with
separations $0.6 \la d \la 1.6$, the `lensing zone.'  The angle
$\alpha$, which is of no physical interest, is uniformly distributed.
Only certain values of $\alpha$ will create detectable deviations.
Thus integration over $\alpha$ defines a geometric detection probability.

\begin{figure}[t]
\plotfiddle{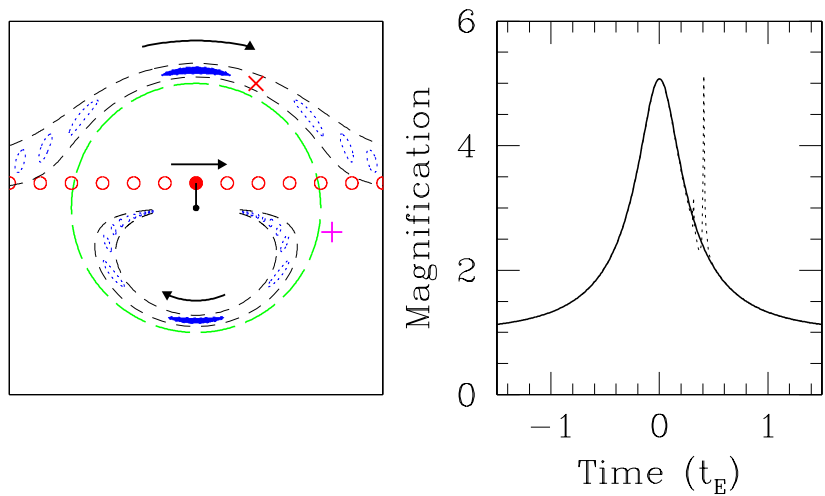}{0.5cm}{0}{70}{70}{-230}{-300}
\vskip6.0cm
\caption{
Left: The images (dotted ovals) are shown for several
different positions of the source (solid circles), along with the primary lens
(dot) and Einstein ring (long dashed circle).  If the
primary lens has a planet near the path of one of the images,
i.e. within the short-dashed lines, then the planet
will perturb the light from the source, creating a
deviation to the single lens light cure.
Right: The magnification as a function of
time is shown for the case of a single lens (solid) and accompanying planet
(dotted) located at the position of the X in the top panel.  
If the planet was located at the + instead, then there would be no detectable perturbation, and
the resulting light curve would be identical to the solid curve.
}
\end{figure}

Microlensing as a method to detect extrasolar planets was first
suggested by Mao \& \Packy (1991), and was expanded upon by Gould \&
Loeb (1992) who demonstrated that if all lenses had a Jupiter analog,
than $\sim 20\%$ of all light curves should exhibit $\ga 5\%$
deviations.  Since these two seminal papers, many authors have
explored the use of microlensing to detect planets.  It is not our
intention to provide a comprehensive review of this field.  However,
of particular relevance is the paper by Griest \& Safizadeh (1998, GS98) who
demonstrated that, for high-magnification events (those with maximum
magnification $\amax > 10$), the detection probability for planets in
the lensing zone is $\sim 100\%$.  Thus high-magnification events are
an extremely efficient means of detecting extrasolar planets. 
The results of GS98 also imply that multiple planets in the lensing zone
should betray their presence in high-magnification events (Gaudi,
Naber \& Sackett 1998). 

\section{Limits on Companions in OGLE-1998-BUL-14}

\begin{figure}[t]
\plottwo{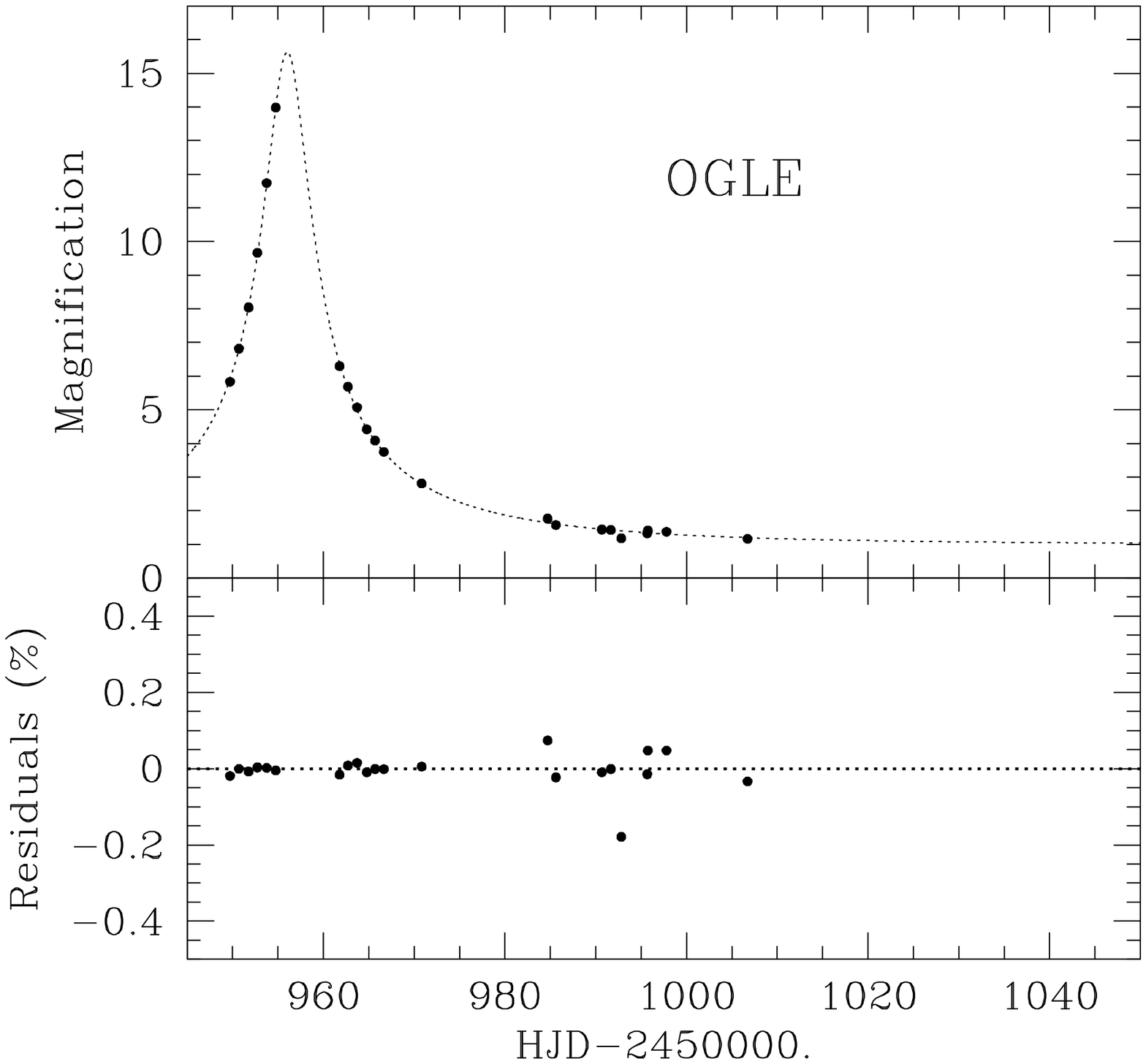}{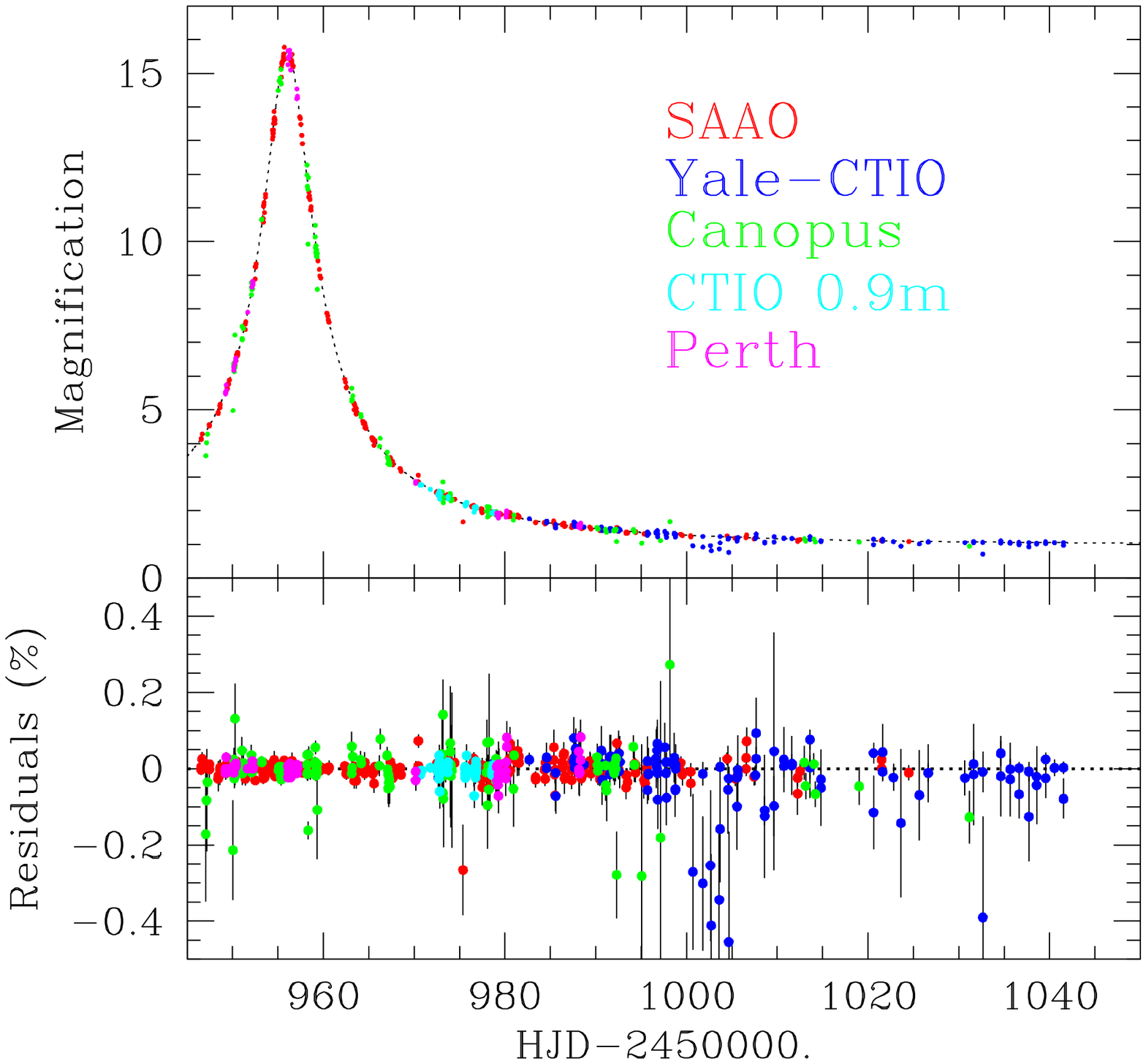}
\caption{
Left: The top panel shows the magnification as a function of time for 
OGLE data of microlensing event OGLE-1998-BUL-14.  The dashed line indicates the best-fit point-source point-lens model (PSPL),  
which has a time scale $\te=40$ days, and a maximum magnification of 
$\sim 16$.  The bottom panel
shows the residuals from the best-fit PSPL model.
Right: The top panel shows the magnification as a function of time for 
PLANET data of microlensing event OGLE-1998-BUL-14.
The bottom panel shows the residuals from the best-fit PSPL model (Albrow et~al.\ 2000).}
\end{figure}

\begin{figure}[t]
\plotfiddle{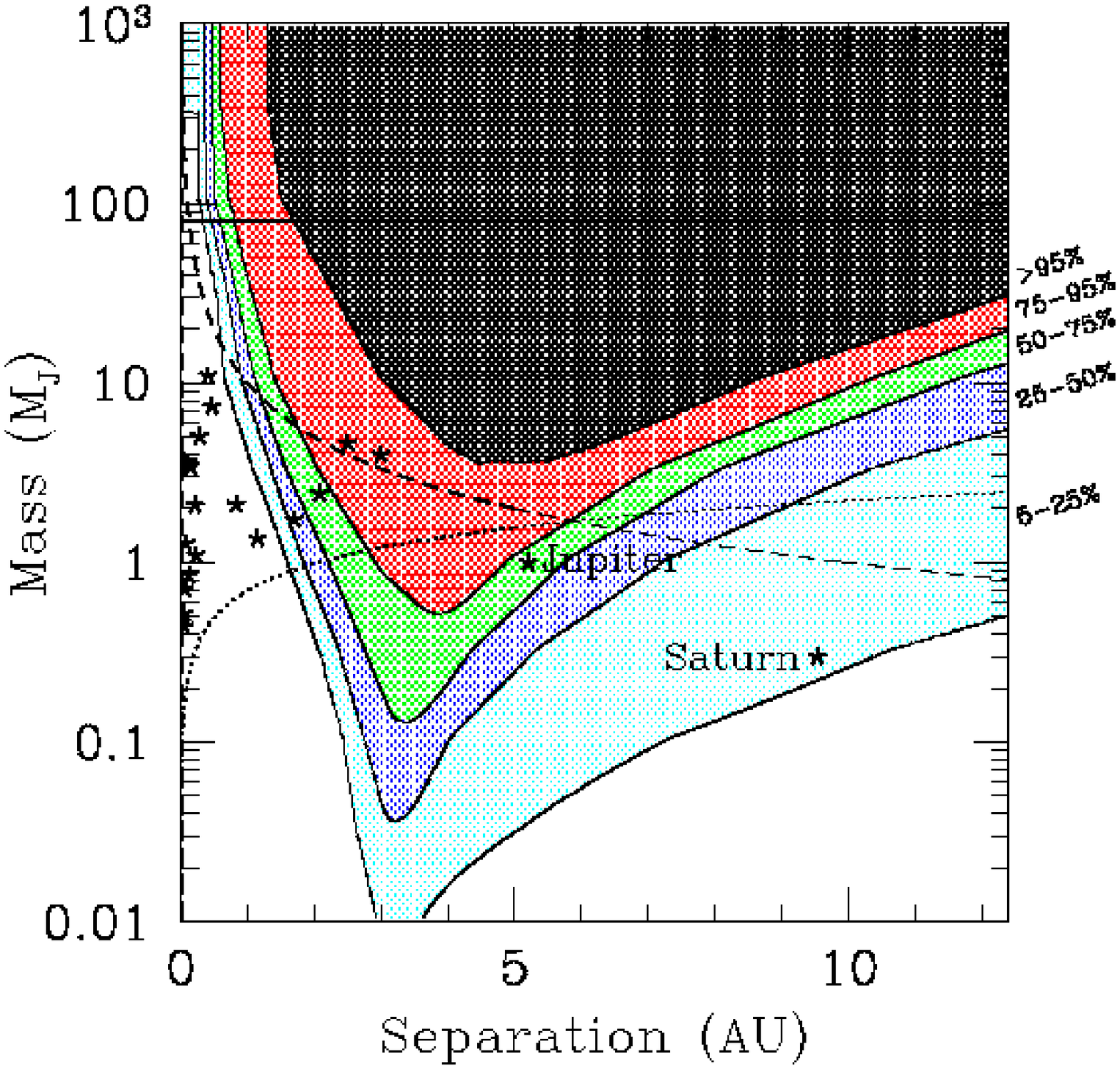}{0.5cm}{0}{50}{40}{-150}{-250}
\vskip7.5cm
\caption{
Detection efficiencies for the PLANET data set of OGLE-1998-BUL-14 as a
function of the mass and orbital separation of
the companion assuming a primary lens mass of $M_\odot$ and Einstein ring radius
 of
$\re=3.1~{\rm AU}$.  The contours are (outer to inner) 5\%,25\%, 50\%,
75\% and 95\%. 
In order to convert from mass ratio and projected
separation to mass and physical separation, we have averaged over
orbital phase and inclination (assuming circular orbits).
Jupiter and Saturn are marked with stars, as are
the extrasolar planets discovered with radial velocity techniques.
The horizontal line marks the hydrogen-burning limit.  The dotted line shows the
radial-velocity detection limit for an accuracy of $20~{\rm m s^{-1}}$
and a primary mass of $M_\odot$.  The dashed line is the
astrometric detection limit for an accuracy of $1~{\rm mas}$ and a
primary of mass $M_\odot$ at $10~{\rm pc}$ (Albrow et~al.\ 2000).   }
\end{figure}

The basic requirements to detect planets with microlensing are good
temporal sampling and photometric precision.  Since the optical depth
to microlensing is low, ${\cal O}(10^{-6})$, the survey
teams that discover microlensing events toward the Galactic bulge must
monitor of order one million stars on a nightly basis in order to detect
any events.  Therefore, they generally have insufficient sampling to
detect the short-duration perturbations to the primary light curve
(see, e.g., Figure 2). However, these survey teams (OGLE, Udalski et~al.\ 1994; MACHO, Alcock
et~al.\ 1996; EROS; Glicenstein et~al., these proceedings) issue
alerts, notification of ongoing events.  This allows follow-up
collaborations (GMAN, Alcock et~al.\ 1997; PLANET, Albrow et~al.\
1998; MPS, Rhie et~al.\ 1999; MOA, Yock, these proceedings) to monitor
these events frequently with high-quality photometry to search for
planetary deviations.  In particular, the PLANET collaboration has
access to four telescopes located in Chile, South Africa, Western
Australia, and Tasmania, and can monitor events nearly
round-the-clock, weather permitting.

Figure 2 shows PLANET photometry of an event alerted by the OGLE
collaboration, \ob14.  This was a high-magnification event ($\amax
\sim 16)$ with $\te\simeq 40~{\rm days}$, making it an excellent
candidate to search for planetary deviations. PLANET obtained a total of
600 data points for this event: 461 $I$-band and 139 $V$-band.  
%Figure 3 demonstrates the quality of the data.  
The median sampling interval is about 1 hour, or $10^{-3}\te$, with
very few gaps greater than 1 day.  The $1\sigma$
scatter in $I$ over the peak of the event (where the sensitivity to
planets is the highest) is $1.5\%$.  The dense sampling and excellent
photometry means that our efficiency to detect massive
companions should be quite high.  In fact, examination of the
residuals from a single lens model (Fig.~2) reveal no obvious
deviations of any kind.

To be more quantitative, we simultaneously search for
binary-lens fits and calculate the detection efficiency
$\epsilon(d,q)$ of \ob14
 as a function of separation and mass ratio using a
method proposed by Gaudi \& Sackett (2000).  For details on the
implementation for this event, see Albrow et~al.\ (2000).  

We find no
binary-lens models in the parameter ranges $0 \le d \le 4$ and
$10^{-5} \le q \le 1$ that provide significantly ($\dchi \ga
10$) better fits to the \ob14 dataset.  We therefore conclude that the
light curve of \ob14 is consistent with a single lens.

In Figure 4 we show the detection efficiency $\epsilon$ of our \ob14
dataset to companions as a function of
the mass $M_{\rm p}$ and orbital radius $a$ of the companion.  Parameter
combinations shaded in black are excluded at the 95\% significance
level.  Stellar companions to the primary lens of \ob14 with 
separations between $\sim 2~{\rm AU}$ and $11~{\rm AU}$ are excluded. 
Companions with mass $\ge 10~M_{\rm J}$ are excluded between $3~{\rm
AU}$ and $7~{\rm AU}$.
 Although we cannot exclude a
Jupiter-mass companion at any separation, we had 
a $\sim 80\%$ chance of detecting such a companion at $3~{\rm AU}$.
The detection efficiency for \ob14 is $>25\%$ at $a=3~{\rm AU}$ for
all companion masses $M_{\rm p}>0.03~M_{\rm J}$.   We find that we had a $\sim 60\%$
chance of detecting a companion with the mass and separation of
Jupiter ($M_{\rm p}=M_{\rm J}$ and $a=5.2~{\rm AU}$),  and a $\sim 5\%$ chance of
detecting a companion with the mass and separation of Saturn ($M_{\rm
p}=0.3~M_{\rm J}$ and $a=9.5~{\rm AU}$) in the light curve of \ob14.

Thus, although Jupiter analogs cannot be ruled out in \ob14, the
detection efficiencies are high enough that non-detections in
several events with similar quality will be sufficient to place
meaningful constraints on their abundance.

How do the \ob14 efficiencies compare to planet detection via other methods?  In Figure~4
we show the radial velocity detection limit on $M_{\rm p}\sin i$ for a
solar mass primary as a
function of the semi-major axis for a velocity amplitude of $K=20~{\rm
m~s^{-1}}$, which is the limit found for the majority of the stars in
the Lick Planet Search (Cumming, Marcy \& Butler 1999).  Although we show this
limit for the full range of $a$, in reality the detection sensitivity 
extends only to $a \la 5~{\rm AU}$ due to the finite
duration of radial-velocity planet searches and the fact that one
needs to observe a significant fraction of an orbital period.
In addition, we plot in Figure~4 the $M_{\rm
p} \sin i $ and $a$ for planetary candidates detected in the Lick survey.
Radial velocity searches clearly probe a different region of parameter
space than microlensing, in particular, smaller separations.  Note, however,
that our \ob14 data set gives us a $> 75\%$ chance of detecting analogs to two
of these extrasolar planets: the third companion to Upsilon
And and the companion to 14 Her.  Although the efficiency
is low, we do have sensitivity to planets with masses as small as
$\sim 0.01 M_{\rm J}$, considerably smaller than can be detected via
radial velocity methods.  For comparison, we also show in Figure~10 the astrometric detection
limit on $M_{\rm p}$ for a $M_\odot$ primary at $10~{\rm pc}$,
for an astrometric accuracy of $\sigma_{\rm A}=1~{\rm mas}$.  For an
astrometric campaign of $11$ years, this limit extends to $\sim
5~{\rm AU}$.  Such an astrometric campaign ($\sigma_{\rm A}=1~{\rm mas}, P=11$ years),
would be sensitive to companions similar to those excluded 
in our analysis of \ob14.   The proposed Space Interferometry Mission (SIM) promises $\sim
4~\mu{\rm as}$ astrometric accuracy, which would permit the detection of
considerably smaller mass companions.

\begin{figure}[t]
\plotfiddle{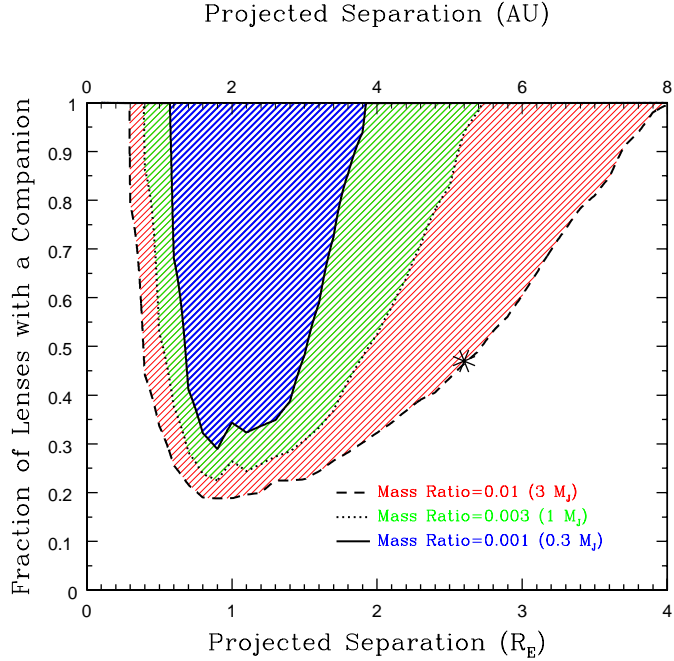}{0.5cm}{0}{50}{50}{-150}{-330}
\vskip9.0cm
\caption{
Upper limits to the fraction of primary lenses with a companion as a
function of the projected separation in $\re$ for three different mass
ratios, $q=10^{-2}$ (dashed), $q=10^{-2.5}$ (dotted) and $q=10^{-3.0}$
(solid).  The projected separation in AU assuming all primary lenses are at $6~\kpc$ and
have masses of $0.3~\msun$ is shown in the top axis.  Similarly, masses
of the secondary are shown in parentheses.  These upper limits are at
the 95\% confidence level and are based on 23 events.}   
\end{figure}

\section{Combined Limits from the 1998-1999 PLANET Seasons}

Clearly one cannot make any statements about the population of the
extrasolar planets as a whole based on one event.  Fortunately, PLANET
has monitored, over the last five years, more than 100 events, a subset of which have
temporal sampling and photometric accuracy similar to that of \ob14.
Here we present a preliminary analysis of these events.

We select 23 high-quality light curves from the 1998-1999 PLANET seasons and analyze
these in the
same manner as \ob14 (Albrow et~al.\ 2000).  Included in this sample
are 5 high-magnification ($\amax > 10$) events: \ob14,
MACHO-1998-BUL-35, OGLE-1999-BLG-5, OGLE-1999-BUL-35 and OGLE-1999-BUL-36.

We find that all 23 events are consistent with a single lens to
within our detection threshold.  Using this null result, along
with the detection efficiency $\epsilon_i(d,q)$ for each event $i$, 
we find a statistical upper limit
to the fraction of these lenses that have a companion of a given
separation and mass ratio.  If the fraction of lenses with a  
companion as a function of $d$ and $q$ is $f(d,q)$, then probability that $N$ events with
individual efficiencies $\epsilon_i(d,q)$ would give a null result (no
detections) is,
\begin{equation}
P = \Pi_{i=1}^{N} [1-f(d,q)\epsilon_i(d,q)].
\end{equation}
The 95\% confidence level (c.l.) upper limit to $f(d,q)$ is found by setting $P= 5\%$.

In Figure 4 we show the 95\% c.l.\ upper limits to $f(d,q)$ for
separations $0\le d \le 4$ and $q=10^{-2},10^{-2.5},$ and $10^{-3}$.
We convert these to limits on the fraction of lenses with companions
of a given mass and physical separation by assuming that all the
primaries have mass $M=\msun$ and distance $\dol=6~\kpc$.  We find that 
$<33\%$ of these lenses have Jupiter mass planets with separations of 1.5-3 AU.  
Similarly, $<33\%$ have planets of mass $M_{\rm p}\ge 3~M_{\rm J}$ with separations of 1-4 AU.  
Although we cannot place an interesting limit on Jupiter analogs, we
do find that $<50\%$ of lenses have $3~M_{\rm J}$ planets at the
separation of Jupiter ($5.2\au$).

\section{Conclusions}

Microlensing offers a unique and complementary method of detecting
extrasolar planets.  Although many light curves have been monitored in
the hopes of detecting the short-duration signature of planetary
companions to the primary lenses, no convincing planetary detections have
yet been made, despite the fact that data of sufficient quality are being
acquired to detect such companions.  These null results
indicate that Jupiter-mass companions with separations in the `lensing
zone,' $1.5-3~\au$, occur is less than 1/3 of systems.  

The potential for this field is enormous.  Current microlensing
searches for planets will continue to monitor events alerted toward the
bulge, and either push these limits down to levels probed by radial
velocity surveys ($\sim 5\%$), or finally detect planets,
and measure the frequency of companions at separations more relevant
to our solar system.  Next generation microlensing
planet searches have the promise of obtaining a robust statistical estimate of the fraction of
stars with planets of mass as low as that of the Earth.   

\acknowledgements
We thank the MACHO, OGLE and EROS collaborations for providing
real-time alerts.  We are especially
grateful to the observatories that support our science (Canopus, CTIO,
Perth and SAAO) via the generous allocations of time that make this
work possible.  This work was supported by grants AST 97-27520 and AST 95-30619 from
the NSF, by grant NAG5-7589 from NASA, by a grant from the Dutch
ASTRON foundation through ASTRON 781.76.018, by a Marie Curie
Fellowship from the European Union, by ``coup de pouce 1999'' award
from Minist{\` e}re de l'{\' E}ducation nationale, de la Rechereche et
de la Technologie, and by a Presidential Fellowship from the Ohio
State University.

\end{document}